\documentclass[reprint,superscriptaddress,amssymb,amsmath,aps,prx,longbibliography, nofo otinbib]{revtex4-2}

\usepackage{graphicx}
\usepackage{amssymb}
\usepackage{epstopdf}
\usepackage{upgreek}
\usepackage{dcolumn}
\usepackage{bm}
\usepackage{textcomp, gensymb}  
\usepackage{braket}
\usepackage{mathtools}
\usepackage{tabularx}
\usepackage{mathrsfs}
\usepackage{lineno}
\usepackage{tikz}
\usepackage{graphicx}
\usepackage{sidecap} 
\usepackage{array}
\usepackage{amsthm}
\usepackage{amsmath}
\usepackage{enumitem}
\usepackage{xcolor} 
\usepackage{amsmath}

\expandafter\ifx\csname package@font\endcsname\relax\else
\expandafter\expandafter
\expandafter\usepackage
\expandafter\expandafter
\expandafter{\csname package@font\endcsname}%
\fi

\newcommand{\be}{\begin{eqnarray}}
\newcommand{\ee}{\end{eqnarray}}
\newcommand{\bfig}{\begin{figure}}
\newcommand{\efig}{\end{figure}}

\DeclareFontFamily{U}{mathb}{}
\DeclareFontShape{U}{mathb}{m}{n}{
  <-5.5> mathb5
  <5.5-6.5> mathb6
  <6.5-7.5> mathb7
  <7.5-8.5> mathb8
  <8.5-9.5> mathb9
  <9.5-11.5> mathb10
  <11.5-> mathbb12
}{}

\usepackage[colorlinks=true,allcolors=blue]{hyperref}%

\begin{document}

\title{Protecting backaction-evading measurements from parametric instability}

\author{E. P. Ruddy}\thanks{These two authors contributed equally}
\email[direct correspondence to ]{yue.jiang@jila.colorado.edu}
\affiliation{JILA, National Institute of Standards and Technology and the University of Colorado, Boulder, Colorado 80309, USA}
\affiliation{Department of Physics, University of Colorado, Boulder, Colorado 80309, USA}

\author{Y. Jiang}\thanks{These two authors contributed equally}
\email[direct correspondence to ]{yue.jiang@jila.colorado.edu}
\affiliation{JILA, National Institute of Standards and Technology and the University of Colorado, Boulder, Colorado 80309, USA}
\affiliation{Department of Physics, University of Colorado, Boulder, Colorado 80309, USA}

\author{N. E. Frattini}
\affiliation{JILA, National Institute of Standards and Technology and the University of Colorado, Boulder, Colorado 80309, USA}
\affiliation{Department of Physics, University of Colorado, Boulder, Colorado 80309, USA}

\author{K. O. Quinlan}
\affiliation{JILA, National Institute of Standards and Technology and the University of Colorado, Boulder, Colorado 80309, USA}
\affiliation{Department of Physics, University of Colorado, Boulder, Colorado 80309, USA}

\author{K. W. Lehnert}
\affiliation{JILA, National Institute of Standards and Technology and the University of Colorado, Boulder, Colorado 80309, USA}
\affiliation{Department of Physics, University of Colorado, Boulder, Colorado 80309, USA}
\date{\today}

\begin{abstract}
Noiseless measurement of a single quadrature in systems of parametrically coupled oscillators is theoretically possible by pumping at the sum and difference frequencies of the two oscillators, realizing a backaction-evading (BAE) scheme. Although this would hold true in the simplest scenario for a system with pure three-wave mixing, implementations of this scheme are hindered by unwanted higher-order parametric processes that destabilize the system and add noise. We show analytically that detuning the two pumps from the sum and difference frequencies can stabilize the system and fully recover the BAE performance, enabling operation at otherwise inaccessible cooperativities. We also show that the acceleration demonstrated in a weak signal detection experiment [PRX QUANTUM 4, 020302 (2023)] was only achievable because of this detuning technique.

\end{abstract}

\maketitle

\section{Introduction}
\label{sec:intro}

Parametric processes such as amplification and frequency-conversion are crucial to the field of quantum information, with broad applications across quantum computing, quantum communication, and quantum sensing \cite{bowen2003experimental,braunstein2005quantum, bergeal2010analog, vijay2011observation, gigan2006self, aspelmeyer2014cavity, backes2021quantum}. Processes that circumvent the quantum noise limit are of particular interest. A well-studied example uses two parametric pumps to evade the quantum backaction induced by measurement \cite{metelmann2015nonreciprocal, chien2020multiparametric, jiang2023accelerated,hertzberg2010back,suh2013optomechanical}. Regrettably, these backaction-evading (BAE) schemes are often thwarted by higher-order nonlinearities which can add noise and cause unstable behavior in the system.

Several strategies have been developed to help mitigate these higher-order effects. For microwave-frequency signals, three-wave mixing Josephson circuits can be designed to have a suppressed fourth-order (Kerr) nonlinearity \cite{frattini20173,sivak2019kerr} at a single operating frequency \cite{chien2020multiparametric} or over a range of frequencies with the addition of an extra flux bias \cite{miano2022frequency}. For measurements of the state of a mechanical oscillator, applying the pumps repeatedly in a pulsed manner allows for the system to relax in between subsequent measurements even in the presence of instability \cite{delaney2019measurement}. Across both the microwave and the optomechanical platforms, strategies that introduce destructive interference using additional pump tones have been proposed but not yet implemented \cite{steinke2013optomechanical, wurtz2021cavity}.

In this article, we present an additional strategy to compensate the dominant fourth-order effect, unwanted single-mode squeezing (SMS), which diminishes BAE performance and leads to instability. By operating with the requisite two parametric pumps detuned from their canonical operation frequencies, we show that BAE performance can be completely recovered even in the presence of this undesired SMS. Crucially, this technique does not require any additional control parameters such as flux biases or pumps, making it simple to implement in an existing experiment and straightforward to combine with any of the aforementioned strategies. To focus our discussion, we describe the detuning technique in the context of one particular application: the microwave amplifier operated with gain and conversion (GC) drives \cite{abdo2013nondegenerate,roy2016introduction,jiang2023accelerated}. 

In Sec.\ \ref{sec:theory}, we discuss the theoretical basis for the technique at the Hamiltonian level. In Sec.\ \ref{sec:characterization}, we characterize the technique for an open quantum system and we demonstrate analytically its effectiveness at recovering BAE performance. Finally, in Sec.\ \ref{sec:sensing}, we describe how the technique can be applied to ultrasensitive quantum sensing experiments with a specific example from a recent BAE search for a microwave-frequency signal \cite{jiang2023accelerated}.

\section{Backaction-evading Hamiltonian Engineering}
\label{sec:theory}

\begin{figure}[t]
	\centering
\includegraphics[width=8.6 cm]{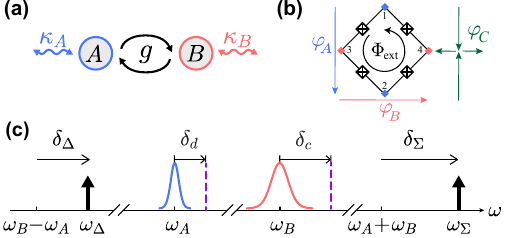} 
	\caption{(a) Simplified mode diagram of a BAE system. (b) Circuit schematic of the JRM. The JRM provides a trilinear interaction term between three modes: two differential modes (blue and red) associated with the north/south and east/west nodes of the bridge respectively, and a common mode (green). (c) Frequency diagram of the modes, pumps, reference frames, and detunings. The pump tones (black arrows) which are applied to the $C$ mode are detuned by $\delta_\Delta$ and $\delta_\Sigma$ with respect to the difference and sum frequencies of the Kerr-shifted science and measurement mode frequencies ($\omega_{A,B}$). The effective Hamiltonian for the system is written in the frame of $\omega_A + \delta_d$ and $\omega_B + \delta_c$ (dashed purple lines).}
	\label{fig:freq_axis}
\end{figure}

The model for the two-tone BAE system, as shown in Fig.\ \ref{fig:freq_axis}(a), consists of a science mode $A$ and a measurement mode $B$ coupled with a state swapping interaction and a two-mode squeezing interaction with matched interaction rates $g$. It is useful to consider an example for how this Hamiltonian may manifest in a physical system such that we may understand how undesired SMS arises. 

To engineer these interactions between microwave-frequency modes, a three-wave mixing Josephson element, such as the Josephson ring modulator (JRM) \cite{bergeal2010analog,bergeal2010phase,abdo2013nondegenerate} may be used. The JRM, depicted schematically in Fig.\ \ref{fig:freq_axis}(b), consists of four identical Josephson junctions with critical current $I_0$ arranged in a ring threaded by an external magnetic flux $\Phi_{\textrm{ext}}$. According to the symmetry of the circuit, three orthogonal normal modes will couple through the JRM: two differential modes $A$ and $B$, and a common mode $C$ \cite{bergeal2010analog,flurin2014josephson}. We note that there exists a fourth electrical mode of the circuit which only contributes to the overall potential of the four nodes uniformly and will therefore be ignored \cite{bergeal2010analog,flurin2014josephson}. Following the conventions in Ref.\ \cite{bergeal2010analog}, the phase differences across the three modes of the JRM, can be expressed as combinations of the fluxes at the four nodes of the circuit $\Phi_{1,2,3,4}$ scaled by the reduced flux quantum $\varphi_0 = \hbar/2e$ as
\begin{equation}
    \varphi_A = \frac{\Phi_1 - \Phi_2}{\varphi_0}; \varphi_B = \frac{\Phi_3 - \Phi_4}{\varphi_0}; \varphi_C = \frac{\Phi_1 + \Phi_2 - \Phi_3 - \Phi_4}{2\varphi_0}.
\end{equation}
The Hamiltonian for the ring is the sum of the Josephson Hamiltonian for each junction \cite{bergeal2010analog}, the details of which are given in Appx.\ \ref{appendix:derivation}. For $\varphi_{A,B,C} \ll 2\pi$, the ring Hamiltonian can be expanded to third order as
\begin{equation}
\begin{aligned}
\label{eq:ring}
    H_{\textrm{ring}}^{(0-3)} = &-4E_J\cos \frac{\varphi_\textrm{ext}}{4}\\
    &+\frac{E_J}{2}\cos \frac{\varphi_\textrm{ext}}{4}\left(\varphi_A^2 + \varphi_B^2 + 4\varphi_C^2 \right)\\
    &-E_J \sin\frac{\varphi_\textrm{ext}}{4} \varphi_A \varphi_B \varphi_C
\end{aligned}
\end{equation}
where $E_J = I_0 \varphi_0$ is the Josephson energy and $\varphi_{\textrm{ext}} = \Phi_{\textrm{ext}}/\varphi_0$. Here, the third-order term is the desired nonlinear coupling term while the terms that are quadratic in $\varphi_{A,B,C}$ only describe the inductive energy stored in the corresponding modes \cite{bergeal2010analog}.

In the idealized conception of the JRM, terms higher than third-order can be neglected as having minimal effect on the dynamics of the system. However, we will see that the next highest order terms (of $\mathcal{O}[\varphi^4]$) destabilize the system and add noise. These harmful terms take the form 
\begin{equation}
\label{eq:fourth_order}
    H_{\textrm{ring}}^{(4)} = -\frac{E_J}{4} \cos{\frac{\varphi_\textrm{ext}}{4}}\left(\varphi_A^2 \varphi_C^2 + \varphi_B^2 \varphi_C^2 +... \right)  
\end{equation}
and mitigating the effects of these terms is precisely the subject of this work. We note that Eq.\ \ref{eq:fourth_order} is given in its complete form in Appx.\ \ref{appendix:derivation}. We begin, however, by studying the dynamics of the idealized system, considering only the terms third-order and lower.

Capacitors are introduced across the nodes \{1,2\} and \{3,4\} of the JRM such that the coordinates $\varphi_{\{A,B,C\}}$ resonate at microwave frequencies. The fields can be quantized by introducing the creation and annihilation operators $a$, $b$, and $c$ \cite{flurin2014josephson} such that $\varphi_A = \frac{\sqrt{Z_A/2}}{\varphi_0} (a + a^{\dag})$, where $Z_A$ is the characteristic impedance of mode $A$ and the others are defined analogously \cite{flurin2014josephson}. Note that here we take $\hbar = 1$.

In the following, we will use the differential $A$ and $B$ modes as the science and measurement modes, and we will use the $C$ mode as the pump mode. For the ideal system in which the only nonlinearity comes from the desired third-order coupling term, the $C$ mode should be pumped at the sum and difference of the $A$ and $B$ mode frequencies ($\omega_\Sigma = \omega_A + \omega_B$ and $\omega_\Delta = \omega_B - \omega_A$) with matched interaction rates $g$ to induce the simultaneous two mode squeezing and state swapping interactions required for the BAE measurement. Because the $C$ mode is driven strongly off its resonance, it can be treated as a classical pump field under the stiff pump approximation \cite{flurin2014josephson}. The interaction Hamiltonian for this system, considering only the terms third-order and lower given in Eq.\ \ref{eq:ring}, is then given by
\begin{equation}
H_{\rm{INT}} =g (A^\dag B + e^{-i\phi} A^\dag B^\dag) +\text{h.c.} 
\label{eq:BAE_fieldoperator}
\end{equation}
where $A$ and $B$ describe the slowly-varying envelopes of the annihilation operators $a$ and $b$ such that $a \rightarrow A e^{-i\omega_A t}$ and $b \rightarrow B e^{-i\omega_B t}$. A more-detailed derivation of the interaction Hamiltonians given in this section can be found in Appx.\ \ref{appendix:derivation}.

We can identify the BAE quadratures of interest by expressing the Hamiltonian in the quadrature basis. Defining the operator for a general quadrature as ${X}_{M,\theta} = \frac{1}{\sqrt{2}} (e^{-i\theta} {M} + e^{i \theta} M^\dag)$ for $M \in \{A, B\}$, we can rewrite Eq.\ \ref{eq:BAE_fieldoperator} as 
\begin{equation}
H_{\rm{INT}} = 2 g X_{A,\phi/2} X_{B,\phi/2}.
\label{eq:Hamiltonian_BAE_quad}
\end{equation}
We see that the phase difference between the microwave drives $\phi$ determines the angle of the amplified quadrature. Without loss of generality, we set $\phi=0$ and define $X_M = X_{M,0}$ and $Y_M = X_{M,\pi/2}$. From Eq.\ \ref{eq:Hamiltonian_BAE_quad}, the Heisenberg equations of motion in the BAE quadratures of interest are given by 
\begin{equation}
    d{Y}_{B}/dt = -2 g {X}_A, \; d{X}_A/dt = 0.
\label{eq: QND EoM}
\end{equation}
These equations indicate that under the ideal BAE interaction, the information contained in $X_A$ appears at $Y_B$ with noiseless amplification and that continuous measurement of $Y_B$ does not perturb $X_A$. Taken together, this implies that the measurement is BAE. In this section, we have focused on a particular system which uses the Josephson nonlinearity to realize a BAE measurement. However, we emphasize that Eq.\ \ref{eq:Hamiltonian_BAE_quad} and the type of BAE measurement it describes can be realized across a variety of systems \cite{aspelmeyer2014cavity,lecocq2015quantum,lei2016quantum}.

Now, we consider how the fourth-order terms given in Eq.\ \ref{eq:fourth_order} modify the equations of motion of the system. The specific choice of drive frequencies causes these terms (which are quadratic in the pump field) to oscillate at $\omega_{\Sigma} + \omega_{\Delta} = 2\omega_B$ and $\omega_{\Sigma} - \omega_{\Delta} = 2\omega_A$, causing parasitic single mode squeezing (SMS). Written in the quadrature basis, these terms modify the interaction Hamiltonian such that
\begin{equation}
H_{\rm{INT}} = 2 g X_A X_B + \frac{s_A}{2} \left(X_A^2-Y_A^2\right)+\frac{s_B}{2} \left(X_B^2-Y_B^2\right)
\label{eq:Hamiltonian_SMS_quad}
\end{equation}
where $s_A$ and $s_B$ are the SMS rates for each mode. Including the SMS terms modifies the equations of motion of motion as
\begin{equation}
\begin{aligned}
    d{Y}_B/dt =& -2 g X_A -s_B X_B, \\
    d{X}_A/dt =& -s_A Y_A, \\
    d{X}_B/dt =& - s_B Y_B, \\
    d{Y}_A/dt =& -2 g X_B - s_A X_A.  
\end{aligned}
\end{equation}
The non-zero time dependence of the signal quadrature $X_A$ signifies that the measurement is not truly BAE, when fourth-order effects are considered.

\begin{figure*}[t]
        \centering
\includegraphics[width=17.2 cm]{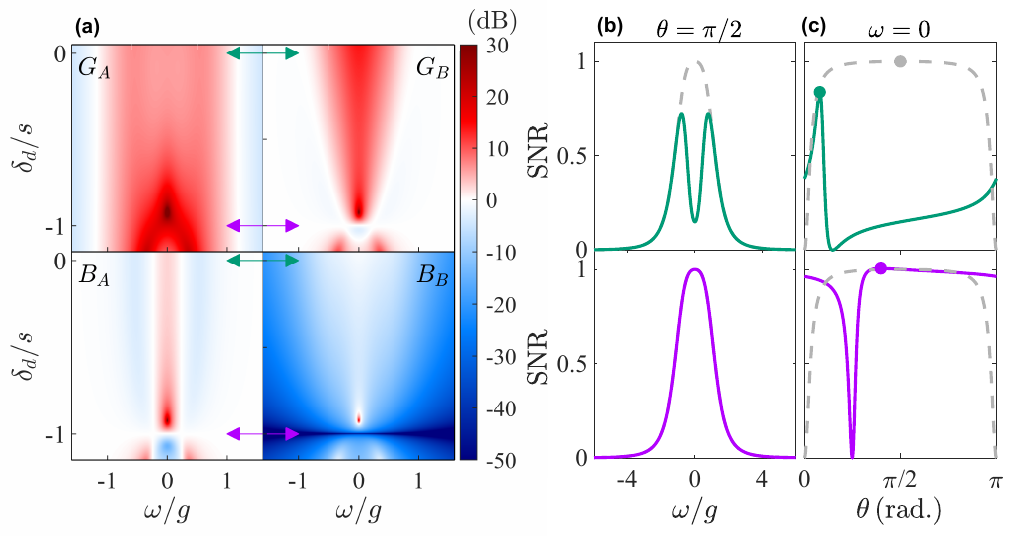}
	\caption{SMS compensation with pump detunings. (a) Gain of the fluctuations entering from ports $A$ and $B$ as measured at $Y_B$, defined analytically in Eqs.\ \ref{eq:scat_paras} - \ref{eq:backactions} for a sample set of conditions. We assume that $s_A = s_B = s$, $\kappa_A = \kappa_B = \kappa$, and $s = g/5 = \kappa/4$. The simulated behavior is broadly tolerant to these choices, as will be explored in Sec.\ \ref{sec:sensing}. The response when no compensatory detunings are applied ($\delta_d = \delta_c = 0$) is marked with a green double-headed arrow and the response when optimal detunings are applied ($\delta_d = -\delta_c = - s$) is marked with a purple arrow. (b) $\textrm{SNR}(\omega)$ for a single-quadrature measurement of $Y_B$, corresponding to a measurement angle of $\theta = \pi/2$. The SNR is normalized to the peak value of the $s = 0$ case (grey, dashed). With no detunings, the peak SNR is reduced due to reduced transmission gain and amplified measurement noise from squeezing. With optimal compensatory detunings, the SNR of the $s = 0$ case is completely recovered, indicating that BAE performance has been restored. (c) $\textrm{SNR}(\theta)$, normalized to its peak value in the $s=0$ case, for a single-quadrature measurement rotated by $\theta$ from $X_B$ at the half-pump frame frequency $\omega = 0$. With or without compensatory detunings, SMS shifts the optimal readout angle away from $\pi/2$ (away from $Y_B$).}
	\label{fig:quad_picture}
\end{figure*}

Fortunately, detuning the applied state swapping and two-mode squeezing drives from the difference and sum frequencies of the $A$ and $B$ modes can cancel the SMS effects on $X_A$ and $Y_B$, depositing all of the undesired effects into the unused quadratures $X_B$ and $Y_A$. We define these intentional detunings $\delta_\Sigma$ and $\delta_\Delta$ such that the applied pump frequencies are given by $\omega_\Sigma = \omega_A + \omega_B + \delta_\Sigma$ and $\omega_\Delta = \omega_B - \omega_A + \delta_\Delta$, as depicted by the horizontal arrows in Fig.\ \ref{fig:freq_axis}(c).

Accounting for the pump detunings, we can express the $a$ and $b$ operators with terms that are slowly varying relative to $(\omega_\Sigma-\omega_\Delta)/2=\omega_A + \delta_d$ for the $A$ mode and $(\omega_\Sigma+\omega_\Delta)/2=\omega_B+\delta_c$ for the $B$ mode. Here, $\delta_d=(\delta_{\Sigma}-\delta_{\Delta})/2$ and $\delta_c=(\delta_{\Sigma}+\delta_{\Delta})/2$  represent the differential and common detunings of the two pumps. Following the steps given in Appx.\ \ref{appendix:derivation}, we arrive at the effective interaction Hamiltonian in the quadrature basis given by
\begin{multline}
H_{\rm{INT}} = 2 g X_A X_B + \frac{s_A}{2} \left(X_A^2-Y_A^2\right)+\frac{s_B}{2} \left(X_B^2-Y_B^2\right) \\ -\frac{\delta_d}{2}\left(X_A^2+Y_A^2\right) -\frac{\delta_c}{2} \left(X_B^2+Y_B^2\right).
\label{eq:Hamiltonian_full_int}
\end{multline}
We find that by carefully choosing $\delta_d = -s_A$, we can compensate the $A$ mode squeezing on the BAE quadratures at the Hamiltonian level. Similarly, choosing $\delta_c = s_B$, we can compensate the $B$ mode squeezing. With these compensatory detunings, the equations of motion for Eq. \ref{eq:Hamiltonian_full_int} become
\begin{equation}
\begin{aligned}
d{Y}_B/dt =& -2 g X_A, \\
d{X}_A/dt =& 0, \\
d{X}_B/dt =& -2 s_B Y_B, \\
d{Y}_A/dt =& -2 g X_B - 2 s_A X_A, \\
\end{aligned}
\label{eq:canceled_eoms}
\end{equation}
where the equations for the BAE quadratures of interest $X_A$ and $Y_B$ have been restored to their ideal forms given in Eq.\ \ref{eq: QND EoM}. Operating at this point deposits all of the undesired effects induced by SMS into the unmeasured quadratures $X_B$ and $Y_A$.

The compensation technique works if the fourth-order terms that give rise to SMS are contained in the Hamiltonian description of the circuit, rather than arising from additional degrees of freedom. This is the case for Josephson circuits where the Hamiltonian takes the form of Eq.\ \ref{eq:appx_FullHam_SMS}, uniquely specifying the terms that result in the desired interactions as well as the higher-order interactions. In optomechanical systems, however, the origin of the SMS is less clear \cite{hertzberg2010back,shomroni2019two}. It is often attributed to a parasitic thermal effect mediated by the mechanical oscillator’s temperature-dependent resonance frequency \cite{suh2012thermally}. Temperature oscillations are caused by the pump power oscillations at twice the mechanical frequency but with a time delay that manifests as a phase shift between the BAE terms and the SMS terms in Eqs.\ \ref{eq:appx_SMS_int_Ham_fieldOp} and \ref{eq:appx_Det_int_Ham_fieldOp}. When the Hamiltonian is expressed in the quadrature basis as in Eqs.\ \ref{eq:Hamiltonian_SMS_quad} and \ref{eq:Hamiltonian_full_int}, the phase shift results in additional terms of the form $X_A Y_A + Y_A X_A$ and $X_B Y_B + Y_B X_B$, precluding the SMS compensation with pump detuning.

\section{Squeezing Compensation in an Open Quantum System}
\label{sec:characterization}

To understand how squeezing compensation would manifest in the presence of noise and loss, we extend our analysis to study an open quantum system. We consider the $A$ mode to be coupled to a signal port at a rate $\kappa_A$, and the measurement mode to be coupled to the readout port at a rate $\kappa_B$, as shown in Fig.\ \ref{fig:freq_axis}(a). We assume for simplicity that both modes have negligible internal loss. We derive the Heisenberg-Langevin equations \cite{clerk2010introduction}
from the full Hamiltonian with generalized detunings and squeezing rates: 
\begin{equation}
\begin{aligned}
\frac{d{Y}_B}{dt} =& -\frac{\kappa_B}{2} Y_B -2 g X_A - (s_B - \delta_c) X_B + \sqrt{\kappa_B} Y_{B,in}, \\
\frac{d{X}_A}{dt} =& -\frac{\kappa_A}{2} X_A - (s_A + \delta_d) Y_A + \sqrt{\kappa_A} X_{A,in}, \\
\frac{d{X}_B}{dt} =& -\frac{\kappa_B}{2} X_B - (s_B + \delta_c) Y_B + \sqrt{\kappa_B} X_{B,in}, \\
\frac{d{Y}_A}{dt} =& -\frac{\kappa_A}{2} Y_A - 2 g X_B - (s_A - \delta_d) X_A + \sqrt{\kappa_A} Y_{A,in}. 
\end{aligned}
\end{equation}

Solving these equations in the frequency domain together with the input-output relations \cite{walls2008quantum} yields the scattering parameters between the $A$ and $B$ ports. We express the scattering parameters in the quadrature basis as a function of $\omega$, the detuning from the half-pump frame frequency. For example, we use $S_{Y_B X_A}$ to denote the output field at the $Y$ quadrature of port $B$ induced by the incoming field at the $X$ quadrature of port $A$. These are given by 
\begin{equation}
\begin{aligned}
S_{Y_B X_A} =& 2g(i \omega + \kappa_A/2)(i\omega + \kappa_B/2) \sqrt{\kappa_A \kappa_B}/\beta,\\
S_{Y_B Y_A} =& -2g(s_A+\delta_d)(i\omega + \kappa_B/2)\sqrt{\kappa_A \kappa_B}/\beta, \\
S_{Y_B X_B} =& \left(4g^2(s_A+\delta_d)+\beta_A(s_B-\delta_c)\right)\kappa_B/\beta, \\
S_{Y_B Y_B} =& 1 - \beta_A(i \omega + \kappa_B/2)\kappa_B/\beta,
\end{aligned}
\label{eq:scat_paras}
\end{equation}
where the quantities
\begin{equation}
\begin{aligned}
\label{eq:betas}
\beta_{A}=\left(i \omega + \kappa_A/2\right)^{2}-\left(s_A^2 - \delta_d^2\right),\\
\beta_{B}=\left(i \omega + \kappa_B/2\right)^{2}-\left(s_B^2 - \delta_c^2\right),\\
\beta = \beta_A\beta_B-4g^2(s_A+\delta_d)(s_B+\delta_c),
\end{aligned}
\end{equation}
have been defined for simplicity.

We can see how compensation restores backaction evasion in the science quadrature of interest, $X_A$, by studying another set of scattering parameters. These are given by
\begin{equation}
\begin{aligned}
S_{X_A X_A} =& 1 - \beta_B(i \omega + \kappa_A/2)\kappa_A/\beta, \\
S_{X_A Y_A} =& \beta_B \left(s_A + \delta_d\right)\kappa_A/\beta,\\
S_{X_A X_B} =& -2g (s_A + \delta_d) (i\omega + \kappa_B/2)\sqrt{\kappa_A \kappa_B}/\beta, \\
S_{X_A Y_B} =& 2g (s_A + \delta_d) (s_B + \delta_c)\sqrt{\kappa_A \kappa_B}/\beta.
\end{aligned}
\label{eq:scat_paras_XA}
\end{equation}

We assume that vacuum fluctuations enter at both ports $A$ and $B$ but that the signal enters only at port $A$. The fluctuations entering at port $B$ therefore represent the noise we wish to evade using the BAE scheme. It is useful to group Eq.\ \ref{eq:scat_paras} into these categories. We define the gain of the fluctuations entering at ports $A$ and $B$ as measured at $Y_B$ as 
\begin{equation}
\begin{aligned}
G_A = |S_{Y_B  X_{A}}|^2+|S_{Y_B Y_{A}}|^2, \\
G_B = |S_{Y_B  X_{B}}|^2+|S_{Y_B Y_B}|^2.
\end{aligned}\label{eq:gains}
\end{equation}
Similarly, the scattering parameters given in Eq.\ \ref{eq:scat_paras_XA} describing the backaction on ${X}_A$ can be categorized according to
\begin{equation}
\begin{aligned}
B_A = |S_{X_A X_{A}}|^2+|S_{X_A Y_{A}}|^2, \\
B_B = |S_{X_A  X_{B}}|^2+|S_{X_A Y_B}|^2.
\end{aligned}\label{eq:backactions}
\end{equation}

In Fig.\ \ref{fig:quad_picture}(a), we plot Eqs.\ \ref{eq:gains} and \ref{eq:backactions} for a sample set of conditions. We assume for simplicity that the modes have equal external coupling rates $\kappa$ and equal squeezing rates $s < \kappa,g$. Note that none of these assumptions are required for compensation. 

Even when the squeezing rates are small compared to $\kappa$, when no compensatory detunings are applied (green double-headed arrow), the amplifier suffers both reduced signal transmission $G_A$ and amplified measurement noise $G_B$ as a result of the squeezing. The poor amplifier performance is a symptom of diminished backaction evasion. This assessment is confirmed by the non-zero transmission of noise $B_B$ from the measurement port to $X_A$, as can be seen in the bottom panel of Fig.\ \ref{fig:quad_picture}(a). In contrast, when optimal compensatory detunings are applied (purple arrow), $B_B(\omega) = 0$, indicating that BAE performance has been restored. It follows, therefore, that under these conditions, the measurement will not suffer amplified measurement noise at $Y_B$. 

Accounting for the vacuum fluctuations entering at both ports $A$ and $B$, the signal-to-noise ratio (SNR) can be found using $\mathrm{SNR} \propto G_A/(G_A + G_B)$. Figure \ref{fig:quad_picture}(b) plots this SNR for the case of no compensatory detunings (green) and optimal detunings (purple), normalized to the ideal ($s = 0$) SNR (grey, dashed). When BAE performance is restored by optimal detunings, the ideal SNR is completely recovered when measuring $Y_B$.

\begin{figure*}
	\centering
	\includegraphics[width=17.2 cm]{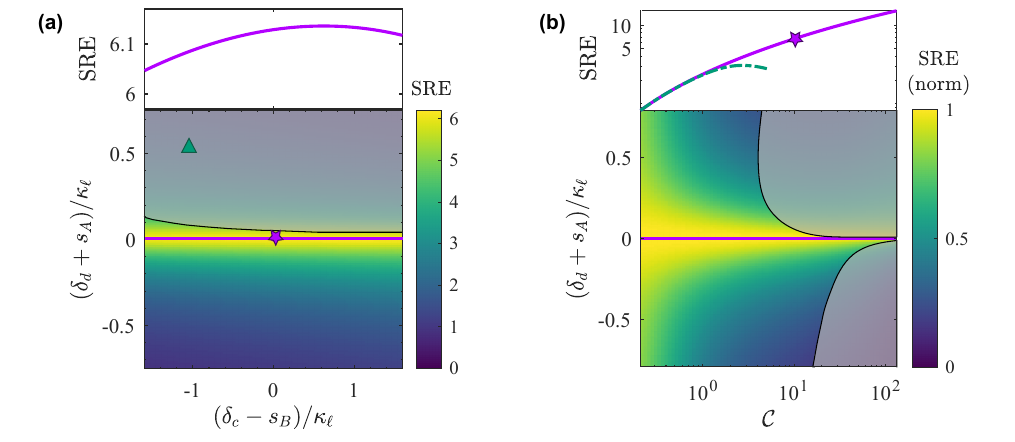} 
	\caption{Scan rate enhancement in the GC-enhanced search for a weak microwave signal at unknown frequency in the presence of unwanted SMS. (a) SRE as a function of pump detunings which are varied about their optimal values ($-s_A$ and $s_B$). Operating points with zero and optimal detunings are marked by the green triangle and purple star. The grey shaded region marks the unstable regime. The condition for canceling the SMS of the science mode satisfies $\delta_d = -s_A$ and is marked by a purple line. The SRE along this line is plotted in the top panel. (b) Robustness of compensation procedure to increased cooperativity $\mathcal{C}$. Upper panel: SRE increases with cooperativity, but uncompensated SMS causes the SRE to turn over (green dashed line) due to amplified measurement noise. The squeezing additionally causes the system to become unstable where the green dashed line ends. Detuning compensation protects the system from these effects, enabling a six-fold acceleration to be achieved and preserving stability as the cooperativity is increased (purple line). Bottom panel: SRE normalized to its optimally compensated value at a given cooperativity. With increased cooperativity, the narrowing of the yellow band reveals an increase in sensitivity to the detuning control parameters. The unstable regime approaches the line of compensation asymptoptically with increased cooperativity (bottom panel), making further SRE beyond $\mathcal{C} = 10.4$ (purple star) difficult to achieve.}
\label{fig:SMS_Cancellation}
\end{figure*}

Crucially, detunings only recover BAE readout along ${Y}_B$. The effects of the squeezing can still be seen when reading out other quadratures. The scattering parameters at $X_B$ are given by 
\begin{equation}
\begin{aligned}
S_{X_B X_A} =& - 2 g (s_B + \delta_c) (i \omega + \kappa_A/2) \sqrt{\kappa_A \kappa_B}/\beta, \\
S_{X_B Y_A} =& 2 g (s_A + \delta_d) (s_B + \delta_c) \sqrt{\kappa_A \kappa_B}/\beta,\\
S_{X_B X_B} =& 1-\beta_A (i\omega 
+ \kappa_B/2) \kappa_B/\beta, \\
S_{X_B Y_B} =& \beta_A (s_B + \delta_c) \kappa_B/\beta.
\end{aligned}
\label{eq:scat_paras_XB}
\end{equation} 
To consider the effects of squeezing on an arbitrary quadrature rotated by $\theta$ from the $X_B$ quadrature, we here perform a rotation on the scattering parameters. We define the rotation matrix 
\begin{equation}
\mathbf{R} = 
\begin{pmatrix}
\cos{\theta} & \sin{\theta} \\
-\sin{\theta} & \cos{\theta}
\end{pmatrix}
\end{equation}
and two input matrices that are subject to rotation
\begin{equation}
\begin{aligned}
    \mathbf{U} = 
    \begin{pmatrix}
        S_{X_B X_A} & S_{X_B Y_A} \\
        S_{Y_B X_A} & S_{Y_B Y_A}
    \end{pmatrix}, 
    \mathbf{W} = 
    \begin{pmatrix}
        S_{X_B X_B} & S_{X_B Y_B} \\
        S_{Y_B X_B} & S_{Y_B Y_B}
    \end{pmatrix}.
\end{aligned}
\end{equation}
Where the scattering parameters as they were first defined are given by $\textbf{S}(\theta = 0) = \left[\mathbf{U},\mathbf{W}\right]$ using Eqs.\ \ref{eq:scat_paras} and \ref{eq:scat_paras_XB}, the rotated scattering parameters $\textbf{S}(\theta)$ are given by
\begin{equation}
\mathbf{S(\theta)} = \left[(\mathbf{R} \mathbf{U} \mathbf{R}^T),
(\mathbf{R} \mathbf{W} \mathbf{R}^T)\right].
\end{equation}
In Fig.\ \ref{fig:quad_picture}(c) we use the rotated scattering parameters $\textbf{S}(\theta)$ to plot the SNR at the half-pump frequency ($\omega = 0$) as a function of the readout quadrature $\theta$. Here, we take $\text{SNR}(\theta) \propto G_A(\theta)/(G_A(\theta) + G_B(\theta))$ where $G_A(\theta) = |S_{X_B X_A}(\theta)|^2 + |S_{X_B Y_A}(\theta)|^2$ and $G_B(\theta)$ is defined analogously.

Uncompensated SMS (green line) causes a reduced SNR in almost all quadratures relative to the $s = 0$ case (grey), and we see that reading out at $\theta = \pi/2$ ($Y_B$) is no longer optimal in this case. Instead, the SNR is maximized along the quadrature marked by the green point. The SNR is maximal here not because the signal has been amplified, but rather because the noise entering at the $B$ port has been squeezed. Notably, the benefit of squeezing requires precise tuning of the phase $\theta$ in contrast to the insensitivity seen in the absence of SMS.

In comparison, when compensatory detunings are applied (purple line), we see that the optimal readout quadrature is largely insensitive to small changes in $\theta$. Additionally, there is actually a modest increase to the peak SNR as compared to the $s = 0$ case. This improvement comes from sacrificing some signal amplification in exchange for squeezed measurement noise along a quadrature marked by the purple dot. In fact, if it were possible to measure along a different quadrature at each frequency $\omega$, the inadvertent squeezing effects could be leveraged to enhance the SNR over the whole measurement range. This could potentially be implemented by applying a frequency-dependent phase shift to the fields exiting the measurement port such that the readout angle is optimized separately for each Fourier component \cite{mcculler2020frequency}.

It should be noted that detuning compensation restores the BAE performance in quadratures defined in the half-pump frame rather than in the mode frame. The measurement's sensitivity will therefore peak for forces driving the $A$ mode at the half-pump frequency $\omega_A + \delta_d$ marked by the purple dashed line in Fig.\ \ref{fig:freq_axis}(c) instead of at $\omega_A$. Fortunately, because the peak sensitivity is fully recovered, as long as the experimentalist is aware of this shift, it can easily be accounted for in experiment. 

In fact, these optimally-detuned BAE operating points can be found naturally in experiments. Mode frequencies are often not known precisely in systems when strong pumps and parametric processes are involved. Tuning up amplifiers therefore usually involves parameter optimization of the pump frequency and power. An optimization procedure designed to search for unit reflections of $G_B$ and transmission gain of $G_A$, as indicated in Fig.\ \ref{fig:quad_picture}, would naturally identify the proper pump frequencies to compensate the undesired effects. However, the performance of the amplifier will only appear to be ideally BAE in a single quadrature. If a two-quadrature measurement is used to tune up the amplifier, the optimal operating point will not appear to be BAE. The squeezing effects should be identifiable by reading out the unprotected quadrature, manifesting as amplified noise. A measurement of this quadrature could therefore give insight into the amplifier's rates of SMS.

\section{Implementation for quantum sensing}
\label{sec:sensing}

In Sec.\ \ref{sec:characterization}, we made the simplifying assumptions that internal loss was negligible compared to the external coupling rates ($\kappa_\ell \ll \kappa_A, \kappa_B$), and that the coupling rates were equal ($\kappa_A = \kappa_B$). While these assumptions generally hold true for amplifiers used in quantum signal processing applications \cite{bergeal2010phase}, this is not usually the case for ultrasensitive quantum sensing applications where the signal is weakly coupled to the science mode \cite{sikivie1985experimental, hertzberg2010back}. In this section, we extend our analysis to consider this quantum sensing application. We treat the system under the assumptions that the internal loss $\kappa_\ell$ of the science mode dominates over its weak coupling to the signal $\kappa_A$, and that both of these rates are small compared to the measurement port coupling rate ($\kappa_A \ll \kappa_\ell \ll \kappa_B$). We maintain the assumption that the internal loss of the measurement mode is negligible. 

In order to focus discussion, we consider one particular sensing application: the search for a weak microwave signal at an unknown frequency, a problem of particular interest for axion dark matter searches \cite{sikivie1985experimental}. Recently, the two-tone BAE measurement technique was applied to an experiment designed to mimic a real axion search, and a six-fold acceleration to the spectral scan rate was demonstrated \cite{jiang2023accelerated}. This acceleration was only possible by successfully canceling undesired SMS using the compensatory detuning scheme described in this paper.

Beyond just destroying the BAE performance, undesired SMS causes these measurement schemes to become unstable, placing a hard limit on the achievable cooperativity. The poles of the scattering parameters (the roots of $\beta$ in Eq.\ \ref{eq:betas}) give the criteria for stability. The existence of a root with a negative imaginary component signifies that the system is unstable. While squeezing destabilizes the system, compensatory detunings counteract the destabilization such that the point of optimal compensation $\delta_d = -s_A$ and $\delta_c = s_B$ is always stable for any combination of the squeezing rates. 

In this section, we analytically predict the scan rate enhancement (SRE) using the system parameters from the axion search demonstration experiment \cite{jiang2023accelerated}. We take $g/2\pi = 7.3 \textrm{ MHz}$, $\kappa_{\ell}/2\pi = 960$ kHz, and $\kappa_{B}/2\pi = 20.6$ MHz. The squeezing rates $s_A$ and $s_B$ are estimated to be around $7\%$ and $14\%$ of the GC interaction rates $g$ respectively. Following the procedure outlined in Sec.\ \ref{sec:characterization}, we calculate the scattering parameters and the SNR for this system. The spectral scan rate scales as $\int \textrm{(SNR)}^{2} d\omega$. The SRE can then be calculated by comparing the scan rate to that of a quantum-limited search. A more detailed discussion of this calculation can be found in Ref.\ \cite{jiang2023accelerated}. 

The SRE for various combinations of pump detunings is given in Fig.\ \ref{fig:SMS_Cancellation}(a), where the unstable regime is represented by the grey shaded area. Because $s_B \ll \kappa_B$ for this system, the squeezing of the $B$ mode has a negligible effect on the system, and it is mainly the $A$ mode squeezing that matters for performance and stability. When no pump detunings are applied ($\delta_c = \delta_d = 0$), the system is already unstable. At the optimal cancellation point, ideal ($s = 0$) amplifier performance is recovered and the system is stabilized. 

Further SRE can theoretically be achieved by pumping harder, corresponding with increased interaction rates $g$ and cooperativity $\mathcal{C} = \frac{4 g^{2}}{\kappa_B (\kappa_\ell+\kappa_A)}$. However, the squeezing terms are quadratic in the pump field $\varphi_C$, whereas the BAE interaction is linear in the pump field. This means that pumping harder increases the squeezing rates relative to both $g$ and the damping rates. We must now consider the robustness of the compensation scheme to increased levels of cooperativity.

In Fig.\ \ref{fig:SMS_Cancellation}(b), the interaction rate $g$ and the squeezing rates $s_A$ and $s_B$ are increased according to their dependence on the pump strength. In the upper panel, we plot the SRE for optimal detunings (purple, $\delta_c=-s_A, \delta_d = s_B$) and zero detunings (green, $\delta_c = \delta_d = 0$). The zero detunings line ends when the system would become unstable. In this case, the maximum SRE that could have been achieved was $\textrm{SRE} = 3.1$ and the system would have become unstable at cooperativity $\mathcal{C} = 5.3$. But in the experimental demonstration, it was possible to achieve $\mathcal{C} = 10.4$, resulting in a scan rate enhancement of $\textrm{SRE} = 5.6$ \cite{jiang2023accelerated}, slightly below the expected value of $6.1$. As discussed in \cite{jiang2023accelerated}, the achieved SRE may have been diminished by drifts in the operating point. 

Although BAE performance and system stability should be achievable to arbitrarily high cooperativity, the requirements on tuning precision become increasingly stringent. In the bottom panel of Fig.\ \ref{fig:SMS_Cancellation}(b), $\delta_c$ is chosen to be optimal ($\delta_c = s_B$) while $\delta_d$ is varied about its optimal value. We see that when $s_A$ becomes large enough, the region of stability vanishes asymptotically with increased cooperativity. In the demonstration experiment, the cooperativity was limited to $\mathcal{C} = 10.4$ for this reason. 

\section{Conclusion and outlook}
\label{sec:conclusion}

The detuning-based technique presented in this article provides a simple strategy for compensating the effects of undesired SMS in two-tone BAE measurements. The compensation scheme could benefit systems both with and without intentional Kerr-suppression while adding very little complexity. This ultimately allows for BAE operation at cooperativities beyond the threshold where destabilizing effects from SMS would have otherwise ruined performance. 

The detuning compensation solution also introduces opportunities to improve on the traditional BAE scheme by further increasing the SNR. By measuring along a slightly different quadrature from the one that is amplified, experiments could implement variational readout \cite{vyatchanin1995quantum,kimble2001conversion,kampel2017improving,mcculler2020frequency}, leveraging the effects of the undesired SMS to achieve both signal amplification and squeezed measurement noise. 

\section*{Acknowledgements} 
\label{section:acknowledgements}
The authors thank John Teufel for useful discussions regarding potential applications of this work to the field of optomechanics. This document was prepared with support from the resources of the Fermi National Accelerator Laboratory (Fermilab), the U.S. Department of Energy, the Office of Science, and the HEP User Facility. Fermilab is managed by Fermi Research Alliance, LLC (FRA), acting under Contract No. DE-AC02-07CH11359 and the NSF award 2209522. Additionally, this work was supported by Q-SEnSE: Quantum Systems through Entangled Science and Engineering (NSF QLCI Award OMA-2016244) and the NSF Physics Frontier Center at JILA (Grant No. PHY-1734006).

\appendix
\section{Interaction Hamiltonian Derivation}
\label{appendix:derivation}

In this section, we provide a detailed derivation of the interaction Hamiltonian given in Eqs.\ \ref{eq:Hamiltonian_BAE_quad}, \ref{eq:Hamiltonian_SMS_quad}, and \ref{eq:Hamiltonian_full_int}, starting from the full JRM Hamiltonian. We begin by recalling that the three modes can be expressed as linear combinations of the fluxes \cite{bergeal2010analog} as
{\begin{equation}
    \varphi_A = \frac{\Phi_1 - \Phi_2}{\varphi_0}; \varphi_B = \frac{\Phi_3 - \Phi_4}{\varphi_0}; \varphi_C = \frac{\Phi_1 + \Phi_2 - \Phi_3 - \Phi_4}{2\varphi_0}.
\end{equation}
The Hamiltonian for the ring is the sum of the Josephson Hamiltonian for each junction, and can be rewritten as a function of the three modes as \cite{bergeal2010analog}
\begin{multline}
\label{eq:appx_Ring_Ham}
    H_{\textrm{ring}} = -4 E_J \left(\cos \frac{\varphi_A}{2} \cos \frac{\varphi_B}{2} \cos \varphi_C \cos \frac{\varphi_{\textrm{ext}}}{4} \right. \\
    \left. + \sin \frac{\varphi_A}{2} \sin \frac{\varphi_B}{2} \sin \varphi_C \sin \frac{\varphi_{\textrm{ext}}}{4} \right) 
\end{multline}
where $E_J = I_0 \varphi_0$ is the Josephson energy and $\varphi_{\textrm{ext}} = \Phi_{\textrm{ext}}/\varphi_0$.}

For $\varphi_{A,B,C} \ll 2\pi$, Eq.\ \ref{eq:appx_Ring_Ham} can be expanded to third order as
\begin{equation}
\begin{aligned}
\label{eq:appx_ring}
    H_{\textrm{ring}}^{(0-3)} = &-4E_J\cos \frac{\varphi_\textrm{ext}}{4}\\
    &+\frac{E_J}{2}\cos \frac{\varphi_\textrm{ext}}{4}\left(\varphi_A^2 + \varphi_B^2 + 4\varphi_C^2 \right)\\
    &-E_J \sin\frac{\varphi_\textrm{ext}}{4} \varphi_A \varphi_B \varphi_C
\end{aligned}
\end{equation}
as given in Sec.\ \ref{sec:theory}. Expanding to one more order ($\mathcal{O}[\Phi^4]$), we can identify the terms that are most harmful to the BAE Hamiltonian and we will see how they produce squeezing and instability. For completeness, here we include all of the fourth-order terms, including the self-Kerr terms. Expanding Eq.\ \ref{eq:appx_ring}, these fourth-order terms are given by 
\begin{equation}
\begin{aligned}
\label{eq:appx_fourth_order}
    H_{\textrm{ring}}^{(4)} =& -E_J \cos{\frac{\varphi_\textrm{ext}}{4}}\left[\frac{\varphi_A^2}{4} \left( \frac{\varphi_A^2}{24} + \frac{\varphi_B^2}{2} + \varphi_C^2 \right) \right. \\ & \left. + \frac{\varphi_B^2}{4}\left( \frac{\varphi_B^2}{24} + \frac{\varphi_A^2}{2} + \varphi_C^2 \right) + \frac{\varphi_C^4}{6} \right].
\end{aligned}
\end{equation}
We will see later that the effect of these fourth-order terms is to shift the resonant frequencies of the modes and that when the $C$ mode is driven at $\omega_{\Delta}$ and $\omega_{\Sigma}$, they additionally induce undesired SMS. But we begin first by considering only the ideal terms which appear at third-order and lower. 

As discussed in Sec.\ \ref{sec:theory}, the coordinates  $\varphi_{\{A,B,C\}}$ resonate at microwave frequencies and we quantize them using the creation and annihilation operators $a$, $b$, and $c$ \cite{flurin2014josephson} such that $\varphi_A = \frac{\sqrt{\hbar Z_A/2}}{\varphi_0} (a + a^{\dag})$ \cite{flurin2014josephson}. The bosonic operators of the three different modes ($a$, $b$ and $c$) commute with each other and the operators associated with the same mode satisfy the usual commutation relation $\left[a, a^{\dag}\right] = 1$.

After quantizing, we can write the system Hamiltonian with its linear and interaction terms as 
\begin{multline}
    H = \omega_A \left(a^{\dag} a + \frac{1}{2}\right) + \omega_B \left(b^{\dag} b + \frac{1}{2}\right) + \omega_C \left(c^{\dag} c + \frac{1}{2}\right) \\ + g_{3} (a + a^{\dag}) (b + b^{\dag}) (c + c^{\dag}) + \mathcal{O}(\Phi^{4})
    \label{eq:appx_FullHamOp}
\end{multline}
where we have set $\hbar = 1$. Here, $g_{3}$ represents the three-wave mixing coupling strength \cite{flurin2014josephson} and $\omega_{\{A,B,C\}}$ describe the frequencies of the three modes. We assume that these frequencies account for all of the shifts that appear when the Hamiltonian is written in normal-ordered form. The Heisenberg equations of motion can be obtained from Eq.\ \ref{eq:appx_FullHamOp} using $\left( \frac{da}{dt} = i \left[H, a \right] \right)$ and these are given by
\begin{equation}
\begin{aligned}
    d{a}/dt =& -i \omega_A a - i g_3 (b + b^{\dag}) (c+ c^{\dag}), \\
    d{b}/dt =& -i \omega_B b - i g_3 (a + a^{\dag}) (c + c^{\dag}).
    \label{eq:appx_EOMs}
\end{aligned}
\end{equation}

For the BAE measurement scheme, the $C$ mode is driven strongly off its resonance frequency at $\omega_{\Delta}$ and $\omega_{\Sigma}$ and can therefore be treated as a classical pump field under the stiff pump approximation \cite{flurin2014josephson}, replacing its annihilation operator with its average value $c \rightarrow \left| \braket{c}\right| \left(e^{-i\omega_{\Delta} t} + e^{-i \omega_{\Sigma}t}\right)$ where we have set the phase difference between the microwave drives $\phi = 0$ without loss of generality. We take $a \rightarrow A e^{-i\omega_A t}$ and $b \rightarrow B e^{-i\omega_B t}$ in order to solve for the behavior of their slowly-varying envelopes $a$ and $b$. Writing the equations of motion for these two modes with the explicit time dependence, treating the $C$ mode as a classical pump field, we find
\begin{equation}
\begin{aligned}
    d{A}/dt =& -i g (B e^{-i\omega_{\Delta} t} + B^{\dag} e^{i\omega_{\Sigma} t}) (e^{-i\omega_{\Delta} t} \\ & + e^{i\omega_{\Delta} t} + e^{-i\omega_{\Sigma} t} + e^{i\omega_{\Sigma} t}), \\
    d{B}/dt =& -i g (A e^{i\omega_{\Delta} t} + A^{\dag} e^{i\omega_{\Sigma} t}) (e^{-i\omega_{\Delta} t} \\ & + e^{i\omega_{\Delta} t} + e^{-i\omega_{\Sigma} t} + e^{i\omega_{\Sigma} t}) 
    \label{eq:appx_PreRWA}
\end{aligned}
\end{equation} 
where we have defined the interaction rate $g = g_3 \left| \braket{c}\right|$. The terms that survive the RWA are given by 
\begin{equation}
\label{eq:appx_postrwa}
\begin{aligned}
    d{A}/dt =& -i g (B + B^{\dag}), \\
    d{B}/dt =& -i g (A + A^{\dag}). 
\end{aligned}
\end{equation}
Equivalently, Eq.\ \ref{eq:appx_postrwa} can be obtained by starting from an effective interaction Hamiltonian of the form 
\begin{equation}
H_{\rm{INT}} =g ( A^\dag  B + A^\dag B^\dag) +\text{h.c.} 
\label{eq:appx_BAE_fieldoperator}
\end{equation}
as given in Eq.\ \ref{eq:BAE_fieldoperator}.

The interaction Hamiltonians given in Eqs.\ \ref{eq:Hamiltonian_SMS_quad} and \ref{eq:Hamiltonian_full_int} can be derived by following the same procedure but including the fourth-order terms and the detuning effects respectively. Accounting for the fourth-order terms, Eq.\ \ref{eq:appx_FullHamOp} becomes 
\begin{multline}
    H = \omega_a \left(a^{\dag} a + \frac{1}{2}\right) + \omega_b \left(b^{\dag} b + \frac{1}{2}\right) + \omega_c \left(c^{\dag} c + \frac{1}{2}\right) \\ + g_{3} (a + a^{\dag}) (b + b^{\dag}) (c + c^{\dag}) + K_{AA} (a + a^{\dag})^{4} \\ + K_{BB} (b + b^{\dag})^{4} + K_{CC} (c + c^{\dag})^{4} + K_{AB} (a + a^{\dag})^{2} (b + b^{\dag})^{2} \\ + K_{AC}(a + a^{\dag})^{2} (c + c^{\dag})^{2} + K_{BC} (b + b^{\dag})^{2} (c + c^{\dag})^{2} 
    \label{eq:appx_FullHam_SMS}
\end{multline}
where we have used $K_{AA}$ to represent the self-Kerr rate of mode $A$ and $K_{AB}$ to represent the cross-Kerr rate between modes $A$ and $B$, defining the others analogously. The Kerr terms will shift the resonance frequencies of the modes and we therefore use $\omega_{a,b,c}$ in Eq.\ \ref{eq:appx_FullHam_SMS} to describe the bare (un-shifted) mode frequencies. The Kerr shifts can be identified by writing the Heisenberg equations of motion for Eq.\ \ref{eq:appx_FullHam_SMS}. For example, the Kerr shift of the $A$ mode will be given by the imaginary coefficient of $a$ in its equation of motion, as shown in Eq.\ \ref{eq:appx_EOMs}. We find that the Kerr-shifted frequencies $\omega_{A,B,C}$ are given by
\begin{equation}
\begin{aligned}
\omega_A &= \omega_a + 12 K_{AA} \left< a^{\dag} a + \frac{1}{2} \right> \\&+ 4 K_{AB} \left< b^{\dag} b + \frac{1}{2}\right> + 4 K_{AC} \left< c^{\dag} c + \frac{1}{2}\right>  \\
\omega_B &= \omega_b  + 12 K_{BB} \left< b^{\dag} b + \frac{1}{2} \right> \\&+ 4 K_{AB} \left< a^{\dag} a + \frac{1}{2}\right> + 4 K_{BC} \left< c^{\dag} c + \frac{1}{2}\right>\\
\omega_C &= \omega_c  + 12 K_{CC} \left< c^{\dag} c + \frac{1}{2} \right> \\&+ 4 K_{AC} \left< a^{\dag} a + \frac{1}{2}\right> + 4 K_{BC} \left< b^{\dag} b + \frac{1}{2}\right>.
\label{eq:appx_kerr}
\end{aligned}
\end{equation}

Using these full expressions for the Kerr-shifted mode frequencies and leaving the mode occupancies unspecified, we accordingly make the substitutions $a \rightarrow A e^{-i\omega_A t}$, $b \rightarrow B e^{-i\omega_B t}$ and $c \rightarrow \left<C\right> e^{-i(\omega_B-\omega_A) t}+ \left<C\right> e^{-i(\omega_B+\omega_A) t}$ into the Heisenberg equations of motion. Under the RWA, the equations of motion can be written as
\begin{equation}
\begin{aligned}
    d{A}/dt =& -i g (B + B^{\dag}) - i s_A A^{\dag}, \\
    d{B}/dt =& -i g (A + A^{\dag}) - i s_B B^{\dag}
\label{eq:appx_SMS_postRWA}
\end{aligned}
\end{equation} 
The effective interaction Hamiltonian for \ref{eq:appx_SMS_postRWA} is then given by
\begin{equation}
H_{\rm{INT}} = g (A^\dag B + A^\dag B^\dag) + s_A A^{\dag 2} + s_B B^{\dag 2} + \text{h.c.} 
\label{eq:appx_SMS_int_Ham_fieldOp}
\end{equation}
where we have defined the single-mode squeezing rates $s_A = 2 K_{AC} \left| \braket{c}\right|^{2}$ and $s_B = 2 K_{BC} \left| \braket{c}\right|^{2}$. We note that other Kerr terms will contribute to these interaction rates and that these will depend on the occupancies of the $A$ and $B$ modes. However, in the regime where the stiff pump approximation is valid, we assume that the pump power is significantly larger than the power of the amplified vacuum, meaning that these contributions are negligible compared to the contributions from the pump. Equation \ref{eq:appx_SMS_int_Ham_fieldOp} is given in its quadrature form in Eq.\ \ref{eq:Hamiltonian_SMS_quad}.

Finally, we return to Eq.\ \ref{eq:appx_FullHam_SMS} to treat our compensatory detunings. Having solved for the Kerr shifts that appear at fourth-order, we again define $\omega_A$ and $\omega_B$ accordingly. However, we now pump the $C$ mode detuned from $\omega_{\Delta}=\omega_B-\omega_A$ and $\omega_{\Sigma}=\omega_B+\omega_A$, taking $c \rightarrow \left| \braket{c}\right| \left(e^{-i (\omega_{\Delta} + \delta_c - \delta_d) t} + e^{-i (\omega_{\Sigma} + \delta_c + \delta_d) t}\right)$. Because of these detunings, it is simplest to solve for the behavior of the envelopes that are slowly varying relative to $(\omega_\Sigma-\omega_\Delta)/2=\omega_A + \delta_d$ for the $A$ mode and $(\omega_\Sigma+\omega_\Delta)/2=\omega_B+\delta_c$ for the $B$ mode. As in Sec.\ \ref{sec:theory}, we therefore take $a \rightarrow Ae^{-i(\omega_A + \delta_d)t}$ and $b \rightarrow Be^{-i(\omega_B + \delta_c)t}$. Making these substitutions, the equations of motion after making the RWA are given by 
\begin{equation}
\begin{aligned}
    d{A}/dt =& -i g (B + B^{\dag}) - i s_A A^{\dag} + i \delta_d A, \\
    d{B}/dt =& -i g (A + A^{\dag}) - i s_B B^{\dag} + i \delta_c B
    \label{eq:appx_SMS_Detuning_postRWA}
\end{aligned}
\end{equation} 
which can be obtained from the effective interaction Hamiltonian 
\begin{equation}
\begin{aligned}
H_{\rm{INT}} &= g (A + A^\dag) (B + B^{\dag}) \\ &+ s_A (A^{\dag 2} + A^2) + s_B (B^{\dag 2}+B^2)
 \\
&- \delta_d \left(A^{\dag} A + \frac{1}{2} \right) - \delta_c \left(B^{\dag} B + \frac{1}{2} \right) 
\label{eq:appx_Det_int_Ham_fieldOp}
\end{aligned}
\end{equation}
which is given in the quadrature basis in Eq.\ \ref{eq:Hamiltonian_full_int}.

\clearpage

\bibliography{main}

\end{document}